\documentclass[conference]{IEEEtran}
\IEEEoverridecommandlockouts
\usepackage{cite}
\usepackage{amsmath,amssymb,amsfonts}
\usepackage{graphicx}
\usepackage{subfig}
\usepackage{url}
\usepackage{tabularx}
\usepackage{booktabs}
\usepackage{booktabs}
\usepackage{siunitx} 
\usepackage{subcaption} 
\def\BibTeX{{\rm B\kern-.05em{\sc i\kern-.025em b}\kern-.08em
    T\kern-.1667em\lower.7ex\hbox{E}\kern-.125emX}}
\begin{document}

\title{Geopolitical Parallax: Beyond Walter Lippmann Just After Large Language Models}

\author{
\IEEEauthorblockN{Mehmet Can Yavuz\IEEEauthorrefmark{1}\IEEEauthorrefmark{2}, Humza Gohar Kabir\IEEEauthorrefmark{1}, and Aylin Özkan\IEEEauthorrefmark{2}}
\IEEEauthorblockA{
\IEEEauthorrefmark{1}\textit{Faculty of Engineering and Natural Sciences, Işık University}, İstanbul, Türkiye \\
\IEEEauthorrefmark{2}\textit{Arky Multimedia}, İstanbul, Türkiye \\
\IEEEauthorrefmark{1}mehmet.yavuz@isikun.edu.tr (Corresponding author)
}
}

\maketitle

\begin{abstract}
Objectivity in journalism has long been contested, oscillating between ideals of neutral, fact-based reporting and the inevitability of subjective framing. With the advent of large language models (LLMs), these tensions are now mediated by algorithmic systems whose training data and design choices may themselves embed cultural or ideological biases. This study investigates \emph{geopolitical parallax}—systematic divergence in news quality and subjectivity assessments—by comparing article-level embeddings from Chinese-origin (Qwen, BGE, Jina) and Western-origin (Snowflake, Granite) model families. We evaluate both on a human-annotated news quality benchmark spanning fifteen stylistic, informational, and affective dimensions, and on parallel corpora covering politically sensitive topics, including Palestine and reciprocal China–United States coverage. Using logistic regression probes and matched-topic evaluation, we quantify per-metric differences in predicted positive-class probabilities between model families. Our findings reveal consistent, non-random divergences aligned with model origin. In Palestine-related coverage, Western models assign higher subjectivity and positive emotion scores, while Chinese models emphasize novelty and descriptiveness. Cross-topic analysis shows asymmetries in structural quality metrics—Chinese-on-US scoring notably lower in fluency, conciseness, technicality, and overall quality—contrasted by higher negative emotion scores. These patterns align with media bias theory and our distinction between semantic, emotional, and relational subjectivity, and extend LLM bias literature by showing that geopolitical framing effects persist in downstream quality assessment tasks. We conclude that LLM-based media evaluation pipelines require cultural calibration to avoid conflating content differences with model-induced bias.
\end{abstract}


\section{Introduction}

The tension between objectivity and subjectivity in journalism dates back to the profession’s earliest days and has been re-kindled by the rise of digital media. At the turn of the twentieth century, “objective journalism” emerged as a methodology that prioritized factual reporting over personal opinion, with practitioners asserting that it is the method—not the journalist—that must remain unbiased. In 1922, Walter Lippmann argued for a social-science approach to reporting, seeking to ground journalism in positivist principles \cite{lippman}. Nearly a century later, the American Press Institute defines bias and objectivity as intertwined concepts: “Journalism attempts to be fair and accurate. It does this through objective methods and managing bias” \cite{apibias}. This modern framing underscores objectivity as a consistent, testable process rather than an absence of perspective.

In contrast, philosophical traditions trace subjectivity back to Descartes’ declaration “I think, therefore I am,” and to Kant’s notion of universal ethics originating in the individual subject. From this vantage, a journalist’s subjective insights can enrich reportage, lending analysis an artistic and philosophical depth \cite{subj}. Yet in practice, subjectivity often overlaps with partisanship: Marxist and Hegelian critiques of the press promoted overtly partisan journalism in service of class struggle. Our focus in this work is on the latter sense of subjectivity—namely, the ideological bias that can shape news narratives.

A well-informed public sphere is essential for democratic discourse. Jürgen Habermas conceptualized the public sphere as a virtual community where citizens engage in reasoned debate \cite{haberbas}. Mass media and social platforms serve as the primary arenas for this discourse, making the ethical codes of journalism foundational to civic life. At the same time, the digital age has ushered in an era of unprecedented news volume: automated content generation and algorithmic aggregation now rival human editors, raising new questions about neutrality and transparency. For instance, in 2018 former U.S. President Donald Trump accused Google of silencing conservative voices, though a statistical study by The Economist found no measurable ideological bias in search results \cite{economist}.

This paper asks: To what extent can we quantitatively measure subjectivity in textual media using computational methods, and how does that measurement shift across geopolitical contexts? We take a model-centric view and compare article-level embeddings from families of large language models trained in Chinese vs Western ecosystems. Using supervised probes for news-quality attributes (e.g., fluency, conciseness, descriptiveness, subjectivity, affect), we score common corpora and analyze alignment and divergence across model families. Then finally, We formalize \emph{geopolitical parallax} as a topic- and metric-conditioned family effect.

\section{Datasets}

In this study, we utilize two complementary datasets to evaluate language model performance across dimensions of quality, ethics, and geopolitical framing. The first is a human-annotated benchmark capturing nuanced dimensions of news article quality. The second is a synthetically generated dataset designed to probe model behavior in politically sensitive and ethically challenging contexts involving China and the United States. Together, these datasets support robust evaluations of textual understanding, bias sensitivity, and ethical generalization.

\begin{table}[h!]
\centering
\caption{Summary Statistics of the News Quality Dataset}
\label{tab:news-quality-summary}
\begin{tabular}{l r}
\toprule
\textbf{Statistic} & \textbf{Value} \\
\midrule
Total Samples & 658 \\
Mean Word Count per Article & 434.03 \\
Standard Deviation of Word Count & 316.05 \\
\bottomrule
\end{tabular}
\end{table}

\subsection{News Quality Dataset}
In order to develop reliable and scalable evaluation methods for news article quality, we use the News Quality Dataset introduced in \cite{arapakis} and the summary statistics given at Table \ref{tab:news-quality-summary}. This dataset addresses editorial control by quantifying a range of textual dimensions and annotator agreement over them. The goal is to identify both global and local features that define high-quality news, ultimately enabling automatic assessment before publication.

The dataset defines several quality dimensions across five broader categories:

\begin{itemize}
    \item \textbf{Readability:} Assessed through \emph{fluency} and \emph{conciseness}. Fluent articles maintain logical and coherent sentence and paragraph transitions. Concise articles avoid redundancy and stay focused on the core topic.
    
    \item \textbf{Informativeness:} This includes \emph{descriptiveness}, \emph{novelty}, \emph{completeness}, and \emph{referencing}. Titles should accurately represent content, introduce new information, cover the topic comprehensively, and reference external sources appropriately.
    
    \item \textbf{Style:} Comprising \emph{formality}, \emph{richness}, and \emph{attractiveness}. Articles should follow formal writing norms, exhibit diverse vocabulary, and feature titles that entice readers without misleading them.
    
    \item \textbf{Topic:} \emph{Technicality} and \emph{popularity} capture how expert-level and widely appealing a topic is.
    
    \item \textbf{Sentiment:} Includes \emph{subjectivity}, \emph{positivity}, and \emph{negativity}, measuring how opinionated, emotionally charged, or sentimentally polarized the content is.
\end{itemize}

\begin{figure}[ht]
    \centering
    \begin{minipage}{0.5\linewidth}
        \centering
        \includegraphics[width=\linewidth]{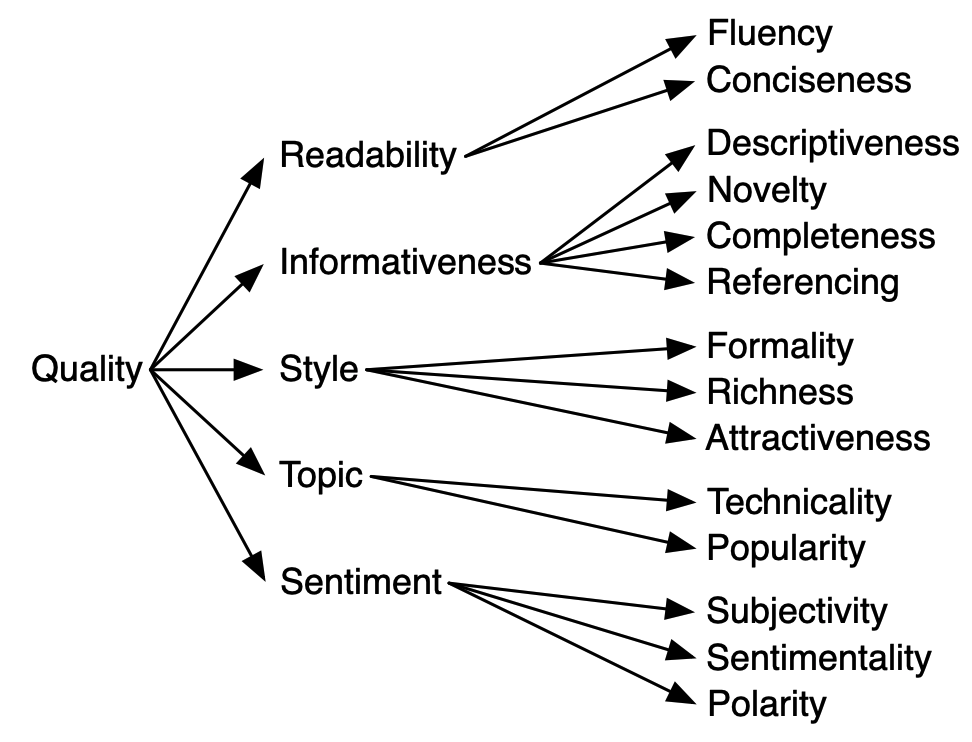}
        \caption{Taxonomy of news article quality dimensions, structured into five main categories: \textit{Readability}, \textit{Informativeness}, \textit{Style}, \textit{Topic}, and \textit{Sentiment}.}
    \end{minipage}
    \hfill
    \begin{minipage}{0.45\linewidth}
        \centering
        \includegraphics[width=\linewidth]{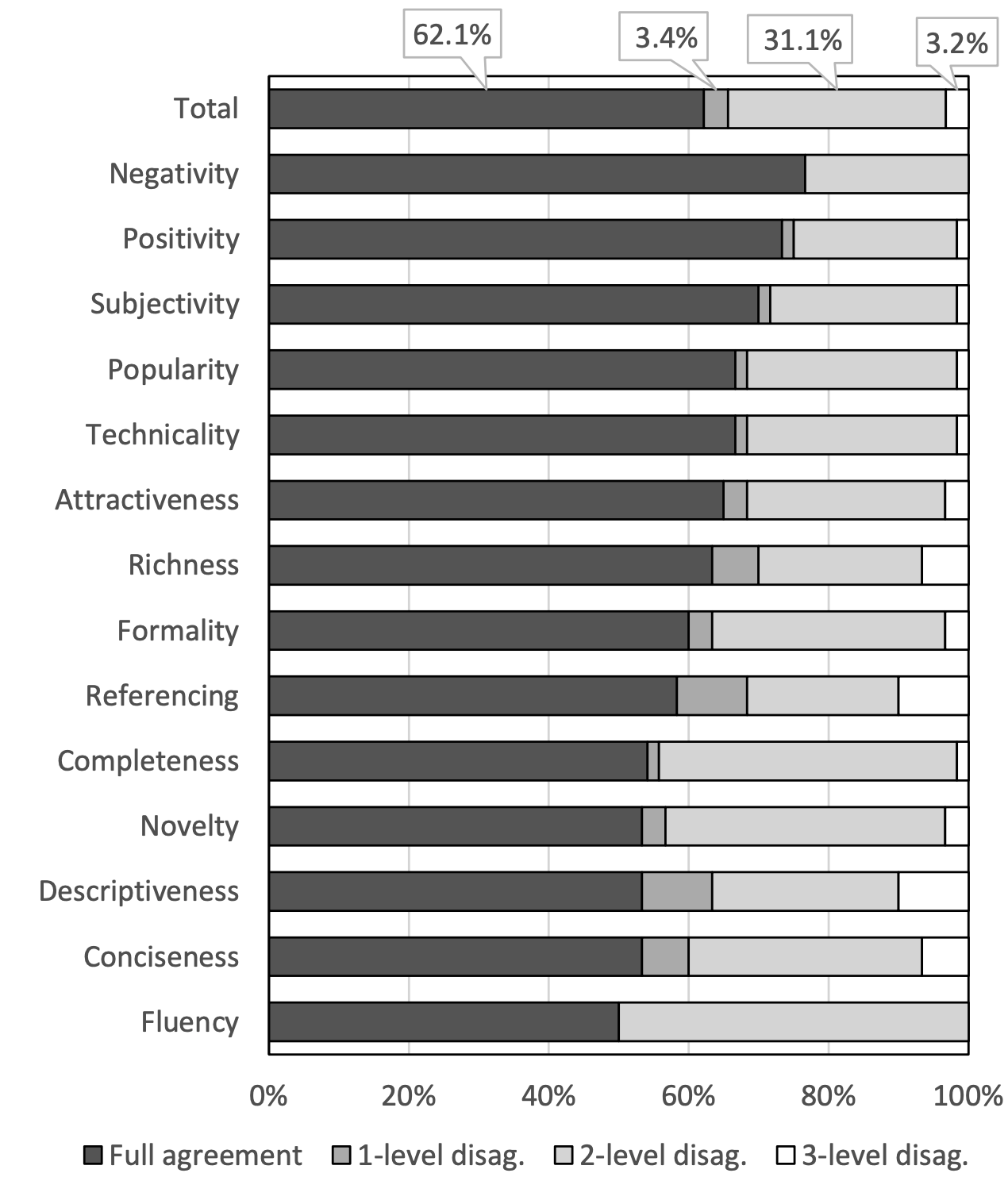}
        \caption{Overview of the News Quality Dataset. (a) Taxonomy of textual quality dimensions. (b) Annotator agreement levels over these dimensions.}    
    \end{minipage}
    \caption{Overview of the News Quality Dataset. (a) Taxonomy of textual quality dimensions. (b) Annotator agreement levels over these dimensions.}
    \label{fig:quality-taxonomy-and-agreement}
\end{figure}

As shown in Figures~\ref{fig:quality-taxonomy-and-agreement}, annotator agreement varies by dimension, with the highest full agreement seen in aspects like \emph{positivity}, \emph{negativity}, and \emph{subjectivity}. More subjective or abstract dimensions such as \emph{fluency}, \emph{novelty}, and \emph{richness} show higher degrees of annotator disagreement. The total full agreement rate is 62.1\%, indicating reasonable inter-annotator reliability for supervised learning applications.

\paragraph{Label Processing}

We use the full set of 15 quality dimensions from the dataset, including \textit{Fluency}, \textit{Conciseness}, \textit{Descriptiveness}, \textit{Novelty}, \textit{Completeness}, \textit{Referencing}, \textit{Formality}, \textit{Richness}, \textit{Attractiveness}, \textit{Technicality}, \textit{Popularity}, \textit{Subjectivity}, \textit{Positive Emotion}, \textit{Negative Emotion}, and overall \textit{Quality}. These continuous-valued annotations are binarized using a threshold of (greater than) 2.0 (on a 1–5 scale), resulting in a balanced binary classification task for each dimension.

\subsection{Generated News for Controversial and Harmful For China vs US Biases}

\begin{table}[h!]
\centering
\caption{Number of Samples per Country with Total Entries}
\label{tab:country_samples_total}
\begin{tabular}{l r}
\toprule
\textbf{Country} & \textbf{Number of Samples} \\
\midrule
China & 1357 \\
Palestine & 1929 \\
US & 1378 \\
\midrule
\multicolumn{2}{c}{\textbf{Total Entries: 4664}} \\
\bottomrule
\end{tabular}
\end{table}

To assess how language models handle politically sensitive and ethically fraught content, we generated a probing dataset comprising synthetic news articles framed around three geopolitical contexts: \textbf{China}, \textbf{the United States} and \textbf{Palestine}. The goal of this dataset is to evaluate the robustness and fairness of embeddings and classification models when exposed to content varying in neutrality, controversy, and harm.

We employed the \texttt{Zephyr-7B-Alpha} model from HuggingFace using a temperature-controlled nucleus sampling strategy (\texttt{top\_p = 0.9}, \texttt{temperature = 0.9}) to promote diverse yet coherent outputs. Each generated paragraph was prompted under one of three controlled framing conditions:

\textbf{Harmful:} Ethically problematic narratives subtly embedding discriminatory or stereotypical framing against \texttt{COUNTRYX}, generated solely for robustness testing in bias and hate speech detection. All harmful outputs are clearly marked and do not reference real entities.

\begin{table*}[h]
\centering
\caption{Examples of Fake News against Different Countries - Research Purpose Only.}
\label{tab:fake_news}
\begin{tabular}{p{3cm} p{10cm}}
\hline
\textbf{Country} & \textbf{Fake News Example} \\
\hline
China & Recently, we have seen evidence of this regime's true nature, as they have taken aggressive action against their own people. The recent protests in Hong Kong have been met with brutal force, as the Chinese government seeks to suppress the peaceful demonstrations of their citizens. \\
\\
Palestine & The Palestinian conflict is not just a local issue. It poses a significant threat to regional stability and international security. The instability and violence in the region have led to the displacement of millions of people and the emergence of new terrorist groups. The situation in Palestine is a breeding ground for radicalism and extremism, and it poses a significant threat to the safety and well-being of people around the world. \\
\\
US & Meanwhile, our political leaders are nothing more than corrupt puppets, controlled by wealthy elites and their own personal interests. They care nothing for the well-being of the people, and are more interested in lining their own pockets with taxpayer money. \\
\hline
\end{tabular}
\end{table*}

Each prompt contained a \texttt{COUNTRYX} placeholder, which we systematically replaced with either \textbf{``China''} or \textbf{``United States''} or \textbf{``Palestine''}to form three parallel datasets shown in Table \ref{tab:country_samples_total} and \ref{tab:fake_news}.

The resulting datasets serve as a diagnostic resource to examine how different embedding models or classifiers respond to subtle variations in narrative framing, region-specific geopolitical bias, and ethical failure modes in language modeling pipelines.These datasets are particularly valuable for evaluating the domain transfer robustness of language representations across different geopolitical framings, their sensitivity to ethically problematic stylistic features, and their capacity for counterfactual generalization when country identifiers are systematically altered.

\section{Methodology}

We use article-level embeddings extracted from two distinct groups of models: Chinese-origin models, including Qwen3, BGE, and Jina embeddings, and Western-origin models, including Snowflake Arctic and IBM Granite embeddings. For each article, the corresponding high-dimensional feature vector is retrieved from each model and subjected to $\ell_2$ normalization to ensure consistency across model scales. Articles are retained for analysis only if they contain both valid embeddings and complete annotations. \footnote{The codebase for the Geopolitical-Parallax project, which explores advanced geopolitical event analysis and visualization, is available at \url{GITHUB: convergedmachine/Geopolitical-Parallax}.}
\begin{table*}[h]
\small
\centering
\caption{Chinese-Origin Embedding Models}
\begin{tabular}{l c c l}
\toprule
\textbf{Model Name} & \textbf{Parameter Size} & \textbf{Feature Vector Length} & \textbf{Abbr.} \\
\midrule
Qwen3-Embedding-8B & 8B & 4096 & Q-8B \\
Qwen3-Embedding-4B & 4B & 2560 & Q-4B \\
Qwen3-Embedding-0.6B & 0.6B & 1024 & Q-0.6B \\
bge-m3 & 3B & 1024 & BGE-M3 \\
bge-large-en-v1.5 & 1.5B & 1024 & BGE-L1.5 \\
bge-base-en-v1.5 & 0.4B & 768 & BGE-B1.5 \\
bge-small-en-v1.5 & 0.1B & 384 & BGE-S1.5 \\
bge-large-en & 0.3B & 1024 & BGE-L \\
bge-base-en & 0.1B & 768 & BGE-B \\
bge-small-en & 15M & 384 & BGE-S \\
jina-embedding-b-en-v1 & 0.4B & 768 & J-B \\
jina-embedding-s-en-v1 & 100M & 512 & J-S \\
\bottomrule
\end{tabular}
\end{table*}

\begin{table*}[h]
\small
\centering
\caption{Western-Origin Embedding Models}
\begin{tabular}{l c c l}
\toprule
\textbf{Model Name} & \textbf{Parameter Size} & \textbf{Feature Vector Length} & \textbf{Abbreviation} \\
\midrule
snowflake-m-v1.5 & 500M & 768 & SF-M1.5 \\
snowflake-l-v2.0 & 1.5B & 1024 & SF-L2.0 \\
snowflake-l & 1.5B & 1024 & SF-L \\
snowflake-m & 500M & 768 & SF-M \\
snowflake-xs & 33M & 384 & SF-XS \\
snowflake-s & 100M & 384 & SF-S \\
granite-125m-en & 125M & 768 & G-125 \\
granite-30m-en & 30M & 384 & G-30 \\
granite-107m-multi & 107M & 384 & G-107M \\
granite-278m-multi & 278M & 768 & G-278M \\
\bottomrule
\end{tabular}
\end{table*}

We use article-level embeddings extracted from two groups of models:
\begin{itemize}
    \item \textbf{Chinese-origin models:} Qwen3, BGE, and Jina embeddings.
    \item \textbf{Western-origin models:} Snowflake Arctic and IBM Granite embeddings.
\end{itemize}

For each article, we retrieve the corresponding high-dimensional feature vector from each model and apply $\ell_2$ normalization to ensure consistency across model scales. Only articles with valid embeddings and annotations are retained.

\subsection{News Quality Assesment}

For each quality label, a logistic regression classifier is trained using the sentence embeddings as input features. The classifier is evaluated under a stratified five-fold cross-validation scheme, which preserves label distributions while ensuring robustness of the evaluation. To address label imbalance within folds, the parameter \texttt{class\_weight="balanced"} is applied. Performance is reported in terms of the weighted F1 score for each label–model pair. The F1 score, defined as the harmonic mean of precision and recall, accounts for both false positives and false negatives while weighting by label support. This training–evaluation cycle is repeated independently for each model across all fifteen quality labels, producing a complete matrix of comparative performance metrics.

The resulting metrics are compiled into two structured tables, one summarizing Chinese-origin models and the other summarizing Western-origin models. These tables are then concatenated into a single dataframe, enabling direct comparison across models and label dimensions. This aggregated comparison facilitates the identification of strengths and weaknesses in the ability of different embedding models to capture linguistic attributes that are relevant to the assessment of news quality.

\subsection{Methodology (II): Geopolitical Parallax Through the Lens of Large Language Models}

This study quantifies cross–model-family divergence in the assessment of subjectivity and related news-quality attributes, a phenomenon we refer to as \emph{geopolitical parallax}. We assembled evaluation sets of news articles associated with four topical–geopolitical pairings: Palestine-related coverage scored by Chinese model families, Palestine-related coverage scored by Western model families, United States–related coverage scored by Chinese model families, and China-related coverage scored by Western model families. For each pairing, the underlying texts were drawn from a shared pool to ensure direct comparability, enabling a matched-topic, cross-model analysis.


Two distinct families of large language model (LLM) embedding generators were selected. The Chinese-origin family comprised Qwen3-Embedding models (0.6B, 4B, 8B), multiple BGE variants (base, large, small, v1.5 releases), BGE-M3, and Jina embeddings. The Western-origin family included Snowflake models of varying sizes and versions, as well as Granite models in both English and multilingual configurations. All article-level embeddings were computed using the models’ default tokenization and pooling strategies, followed by L2 normalization to remove scale effects.

We operationalized subjectivity and related quality constructs using a predefined label set spanning semantic quality (fluency, conciseness, descriptiveness, technicality, novelty), emotional tone (positive emotion, negative emotion), and relational or perceptual qualities (subjectivity, attractiveness, popularity, overall quality). For each model family, binary logistic regression probes were trained to predict the presence of each label from embeddings. Training employed stratified five-fold cross-validation to maintain balanced label distributions across folds. Logistic regression was implemented with L2 regularization and the \texttt{liblinear} solver, with hyperparameters set to scikit-learn defaults unless otherwise specified.

For each topic, model family, and label, we computed the mean predicted positive-class probability over all evaluated sentences. Scores were then averaged across all models within a family to produce a family-level mean for each label, thereby reducing model-specific variance. To quantify divergence between families, we calculated the difference in family-level means, the \emph{parallax delta}, is defined as
\[
\Delta_{t,l} = \bar{p}_{\text{Chinese},t,l} - \bar{p}_{\text{Western},t,l}
\]
where $\bar{p}_{\text{Chinese},t,l}$ and $\bar{p}_{\text{Western},t,l}$ are the average positive-class probabilities for label $l$ on topic $t$ as predicted by the Chinese and Western model families, respectively. Positive values of $\Delta_{t,l}$ indicate higher scores from Chinese models, while negative values indicate higher scores from Western models. This delta computation was applied both for matched topics (e.g., Palestine: Chinese vs.\ Western) and for cross-topics (Chinese-on-US vs.\ Western-on-China), allowing detection of asymmetries in the evaluation of each other’s national coverage.

\section{Results (I): News Quality Assesment}

We evaluate the performance of both Chinese-origin and Western-origin embedding models on 15 news quality dimensions using binary classification tasks. Each embedding is fed into a logistic regression classifier, and weighted F1 scores are computed via 5-fold stratified cross-validation. Tables~\ref{tab:chinese_models} and~\ref{tab:western_models} summarize the results across all models and quality labels.

\subsection{Chinese-Origin Models}

As shown in Table~\ref{tab:chinese_models}, several models from the Chinese-origin family perform competitively across different quality dimensions. \texttt{bge-large-en} achieves the highest F1 scores on \textit{Completeness} (0.764) and \textit{Quality} (0.769), while \texttt{bge-m3} performs best on \textit{Attractiveness} (0.706), \textit{Referencing} (0.663), and ties for top score in \textit{Richness} (0.762). \texttt{Qwen3-Embedding-8B} and \texttt{Qwen3-Embedding-4B} consistently rank among the top three performers in several dimensions, such as \textit{Fluency}, \textit{Descriptiveness}, and \textit{Positive Emotion}. Notably, \texttt{jina-embedding-b-en-v1} achieves the highest F1 in \textit{Negative Emotion} (0.800), indicating strong sentiment modeling capabilities.

\subsection{Western-Origin Models}

Table~\ref{tab:western_models} presents results for Western-origin models, where the \texttt{Snowflake-arctic-embed} series generally demonstrates superior performance. In particular, \texttt{snowflake-m-v1.5} leads in \textit{Conciseness} (0.742) and \textit{Novelty} (0.705), while \texttt{snowflake-l-v2.0} achieves the highest score in \textit{Subjectivity} (0.785). These results suggest that Snowflake models are especially effective at modeling stylistic and structural aspects of text. The IBM Granite models perform comparably in some cases, especially in \textit{Conciseness}, \textit{Descriptiveness}, and \textit{Positive Emotion}, but generally rank lower than their Snowflake counterparts.

\subsection{Comparative Trends}

Overall, both origin groups exhibit strong performance across different subsets of quality labels. Chinese-origin models tend to dominate in content-focused dimensions such as \textit{Completeness}, \textit{Descriptiveness}, and \textit{Referencing}, while Western-origin models, particularly the Snowflake family, demonstrate strengths in style-related labels such as \textit{Conciseness}, \textit{Novelty}, and \textit{Subjectivity}. This highlights a possible divergence in model pretraining objectives or linguistic generalization, reflecting region-specific language modeling priorities.

To further aid interpretability, the best-performing model for each label is \textbf{bolded}, and the second and third best are \underline{underlined} in both tables.

\begin{table*}[htbp]
\centering
\scriptsize
\resizebox{\textwidth}{!}{%
\begin{tabular}{lllllllllllll}
\toprule
           Label & Q-8B & Q-4B & Q-0.6B &            BGE-M3 &          BGE-L1.5 &          BGE-B1.5 &          BGE-S1.5 &BGE-L &BGE-B &BGE-S &  J-B &   J-S \\
\midrule
  Attractiveness &0.661 & \underline{0.687} &  0.659 &    \textbf{0.706} &0.660 &0.652 &0.630 &0.664 &0.662 &0.610 &0.625 & 0.632 \\
    Completeness &0.711 &0.741 &  0.725 & \underline{0.761} &0.726 & \underline{0.761} &0.671 &0.724 &    \textbf{0.764} &0.640 &0.665 & 0.681 \\
     Conciseness &0.723 &0.726 &  0.714 & \underline{0.734} &0.722 &0.716 &0.707 &0.719 &0.726 &0.688 &0.689 & 0.699 \\
 Descriptiveness & \underline{0.694} &    \textbf{0.695} &  0.654 & \underline{0.692} &0.660 &0.636 &0.662 &0.674 &0.644 &0.646 &0.676 & 0.664 \\
         Fluency &    \textbf{0.723} & \underline{0.718} &  0.713 & \underline{0.716} &0.687 &0.690 &0.658 &0.678 &0.679 &0.635 &0.660 & 0.662 \\
       Formality &0.587 &0.589 &  0.579 & \underline{0.624} & \underline{0.611} &0.587 &0.593 &0.600 &    \textbf{0.628} &0.584 &0.588 & 0.564 \\
Negative Emotion &0.788 &0.786 &  0.769 &0.768 &0.788 &0.791 & \underline{0.795} &0.785 &0.784 & \underline{0.800} &    \textbf{0.800} & 0.782 \\
         Novelty &0.690 &0.683 &  0.671 & \underline{0.699} &0.660 &0.680 &0.679 &0.652 &0.654 &0.664 &0.668 & 0.648 \\
      Popularity & \underline{0.679} &0.670 &  0.648 &    \textbf{0.680} &0.625 &0.656 &0.646 &0.653 &0.661 &0.632 &0.662 & 0.640 \\
Positive Emotion &    \textbf{0.821} &0.799 &  0.809 &0.805 & \underline{0.816} &0.798 &0.803 & \underline{0.813} &0.787 &0.803 &0.776 & 0.793 \\
         Quality &0.728 &0.748 &  0.728 & \underline{0.766} &0.716 & \underline{0.750} &0.669 &0.723 &    \textbf{0.769} &0.644 &0.676 & 0.689 \\
     Referencing &0.609 &0.600 &  0.594 &    \textbf{0.663} &0.632 & \underline{0.643} &0.629 &0.636 &0.639 &0.620 &0.599 & 0.579 \\
        Richness &0.703 &0.733 &  0.704 & \underline{0.762} &0.715 &    \textbf{0.762} &0.676 &0.718 & \underline{0.749} &0.657 &0.682 & 0.681 \\
    Subjectivity & \underline{0.784} &0.767 &  0.762 & \underline{0.779} &0.777 &0.760 &0.754 &0.774 &0.766 &0.749 &0.768 & 0.757 \\
    Technicality &    \textbf{0.697} &0.678 &  0.688 &0.665 &0.681 &0.677 &0.682 & \underline{0.695} &0.682 &0.669 & \underline{0.696} & 0.684 \\
\bottomrule
\end{tabular}

}
\caption{Chinese-origin models. Best values are \textbf{bolded}, second and third best are \underline{underlined}.}
\label{tab:chinese_models}
\end{table*}

\begin{table*}[htbp]
\centering
\scriptsize
\resizebox{\textwidth}{!}{%
\begin{tabular}{lllllllllll}
\toprule
           Label & SF-M1.5 &SF-L2.0 & SF-L & SF-M & SF-XS &  SF-S & G-125 &  G-30 & G-107M & G-278M \\
\midrule
  Attractiveness &   0.661 &0.668 &0.641 & \underline{0.670} & 0.644 & 0.646 & 0.639 & 0.660 &  0.622 &  0.623 \\
    Completeness &   0.709 &0.726 &0.688 &0.707 & 0.718 & 0.685 & 0.642 & 0.678 &  0.651 &  0.632 \\
     Conciseness &   0.713 & \underline{0.729} &0.728 &    \textbf{0.742} & 0.697 & 0.689 & 0.678 & 0.724 &  0.700 &  0.698 \\
 Descriptiveness &   0.672 &0.677 &0.691 &0.683 & 0.661 & 0.664 & 0.680 & 0.676 &  0.658 &  0.661 \\
         Fluency &   0.703 &0.704 &0.681 &0.679 & 0.709 & 0.677 & 0.648 & 0.676 &  0.676 &  0.662 \\
       Formality &   0.605 &0.590 &0.567 &0.577 & 0.575 & 0.571 & 0.561 & 0.572 &  0.599 &  0.576 \\
Negative Emotion &   0.794 &0.789 &0.773 &0.787 & 0.780 & 0.778 & 0.769 & 0.768 &  0.771 &  0.761 \\
         Novelty &   0.678 &0.688 & \underline{0.694} &    \textbf{0.705} & 0.663 & 0.692 & 0.663 & 0.657 &  0.654 &  0.658 \\
      Popularity &   0.661 &0.666 &0.659 & \underline{0.673} & 0.672 & 0.658 & 0.643 & 0.668 &  0.637 &  0.632 \\
Positive Emotion &   0.793 &0.807 &0.798 &0.801 & 0.799 & 0.783 & 0.789 & 0.786 &  0.793 &  0.788 \\
         Quality &   0.714 &0.740 &0.702 &0.726 & 0.710 & 0.705 & 0.665 & 0.729 &  0.641 &  0.642 \\
     Referencing &   0.617 & \underline{0.645} &0.596 &0.610 & 0.636 & 0.589 & 0.588 & 0.623 &  0.575 &  0.578 \\
        Richness &   0.695 &0.712 &0.682 &0.714 & 0.721 & 0.693 & 0.636 & 0.699 &  0.648 &  0.636 \\
    Subjectivity &   0.753 &    \textbf{0.785} &0.762 &0.753 & 0.760 & 0.772 & 0.746 & 0.754 &  0.741 &  0.757 \\
    Technicality &   0.676 &0.665 &0.683 &0.673 & 0.669 & 0.659 & 0.666 & 0.667 &  0.672 &  0.671 \\
\bottomrule
\end{tabular}

}
\caption{Western-origin models . Best values are \textbf{bolded}, second and third best are \underline{underlined}.}
\label{tab:western_models}
\end{table*}

\section{Results (II): Evaluating Geopolitical Parallax}

\begin{figure*}[t]
    \centering
    \begin{minipage}[t]{0.48\textwidth}
        \centering
        \includegraphics[width=\linewidth]{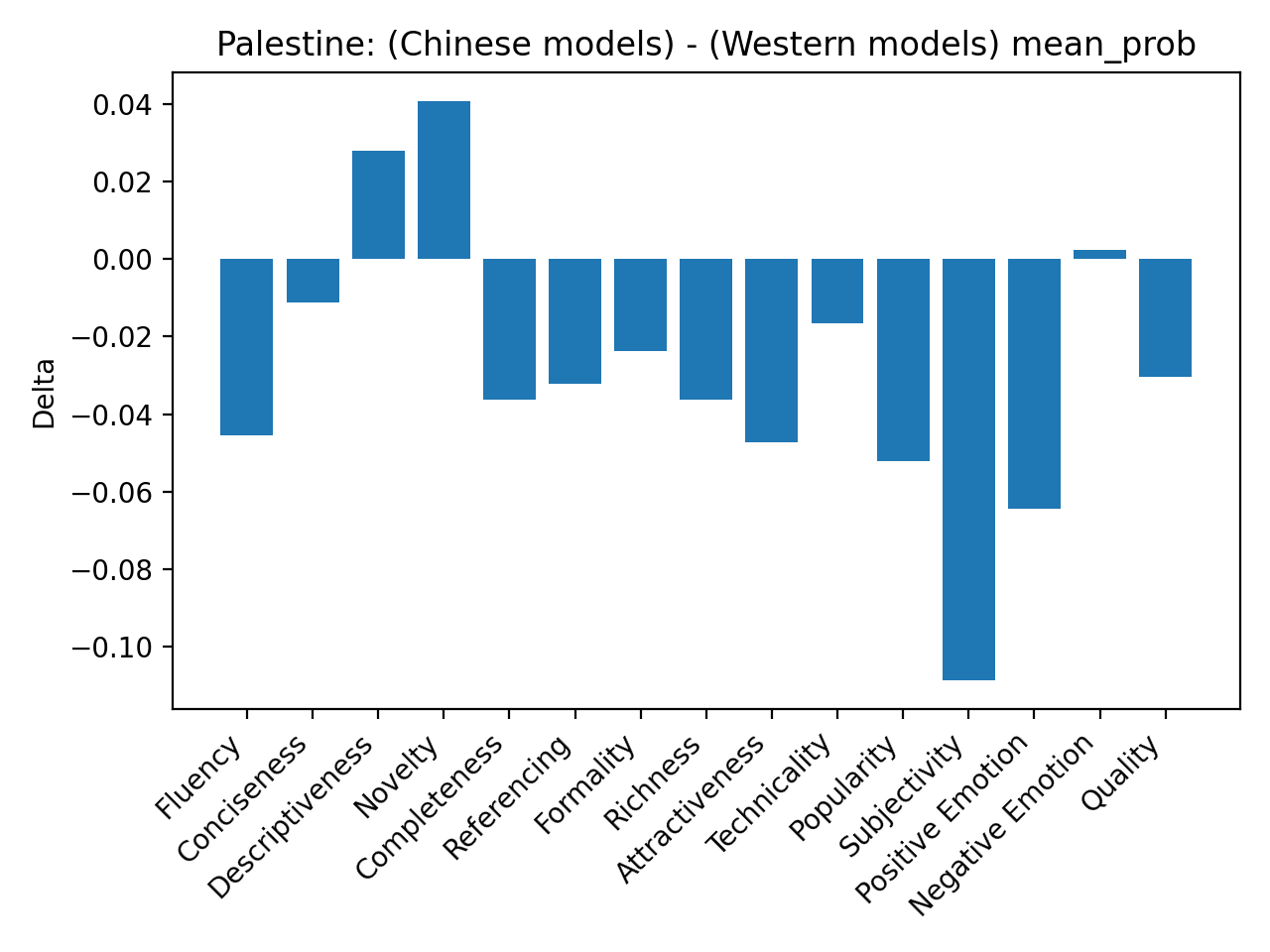}
        \caption*{(a) Metric-level differences in mean predicted probability for Palestine-related coverage: (Chinese models) $-$ (Western models). Positive values indicate higher scores from Chinese models.}
        \label{fig:palestine_delta}
    \end{minipage}
    \hfill
    \begin{minipage}[t]{0.48\textwidth}
        \centering
        \includegraphics[width=\linewidth]{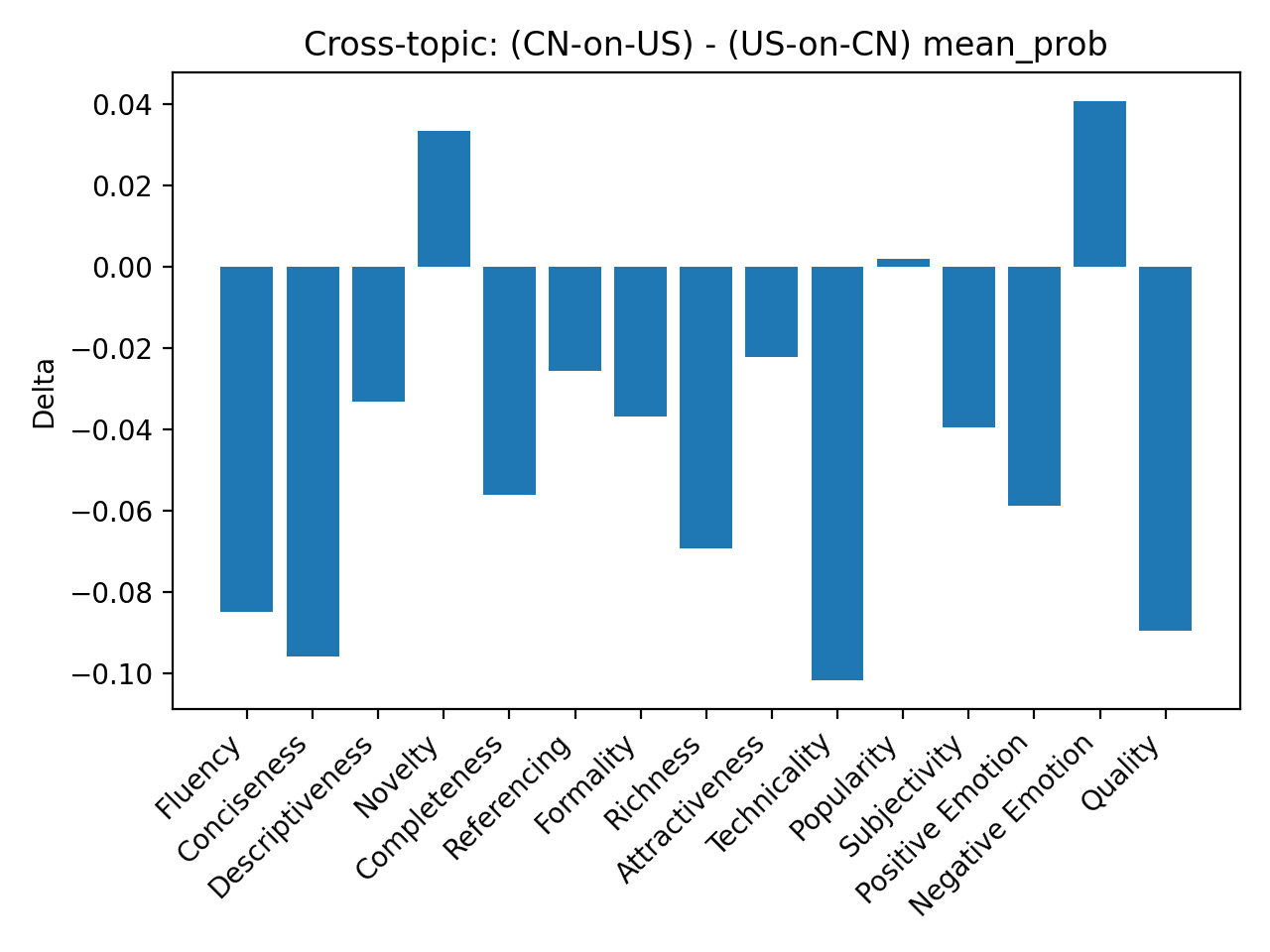}
        \caption*{(b) Metric-level differences in mean predicted probability for cross-topic evaluation: (Chinese-on-US) $-$ (Western-on-China). Positive values indicate higher scores from Chinese models.}
        \label{fig:crosstopic_delta}
    \end{minipage}
    \caption{Comparison of per-metric score \textbf{Parallax deltas ($\Delta$)} between Chinese and Western model families for (a) Palestine-related coverage and (b) cross-topic evaluation.}
    \label{fig:combined_deltas}
\end{figure*}

Figures ~\ref{fig:combined_deltas} present the per-metric differences in predicted positive-class probability between Chinese and Western model families for the Palestine and cross-topic settings, respectively. In both plots, the vertical axis represents $\Delta$ (Chinese minus Western), with positive values indicating higher scores from the Chinese models.

For the Palestine-related corpus, the mean overall difference across all metrics was approximately $-0.029$, indicating that Western models tended to assign higher scores. The largest divergence occurred for \emph{Subjectivity} ($\Delta \approx -0.109$), where Western models rated Palestine coverage as substantially more subjective. Similar negative shifts were observed for \emph{Positive Emotion} ($\Delta \approx -0.064$), \emph{Fluency} ($\Delta \approx -0.046$), \emph{Attractiveness} ($\Delta \approx -0.052$), and \emph{Popularity} ($\Delta \approx -0.050$). In contrast, Chinese models assigned higher scores for \emph{Novelty} ($\Delta \approx +0.041$) and \emph{Descriptiveness} ($\Delta \approx +0.028$), suggesting a relative emphasis on informational uniqueness and richness of description.

In the cross-topic analysis, where Chinese models evaluated United States–related coverage and Western models evaluated China-related coverage, the asymmetry was more pronounced, with an average overall $\Delta$ of approximately $-0.043$. Here, the strongest negative differences were found for \emph{Technicality} ($\Delta \approx -0.102$), \emph{Conciseness} ($\Delta \approx -0.096$), \emph{Quality} ($\Delta \approx -0.090$), and \emph{Fluency} ($\Delta \approx -0.085$), indicating consistently lower evaluations by Chinese-on-US compared to Western-on-China. The only notable positive divergence was in \emph{Negative Emotion} ($\Delta \approx +0.041$), suggesting a greater perceived negative affect in US-related coverage as scored by the Chinese models.

These results reveal that model family origin systematically influences downstream quality assessments in a manner aligned with geopolitical context. In the Palestine case, differences align with affective tone versus descriptive novelty, while the cross-topic results point to broader asymmetries in the appraisal of linguistic and structural quality.

\section{Discussion}

The results demonstrate that large language model embeddings, when used as the basis for quality and subjectivity assessments, carry systematic differences that align with the geopolitical origin of the model family. This finding is consistent with longstanding observations in media studies that framing, tone, and evaluative criteria are not purely universal but are shaped by institutional, cultural, and political contexts. In the Palestine-related analysis, Western models produced higher scores for subjectivity and positive emotion, implying that coverage of this topic, as represented in the shared corpus, is interpreted through a lens that foregrounds affective and evaluative framing. Chinese models, by contrast, emphasized novelty and descriptiveness, suggesting a greater weight placed on informational differentiation and narrative detail.

From the perspective of bias theory, these divergences can be read as differences in \emph{semantic subjectivity}—where model outputs reflect culturally conditioned preferences for certain lexical or narrative features—and \emph{emotional subjectivity}, where the intensity or valence of sentiment is variably amplified. The cross-topic results amplify this interpretation: when scoring politically sensitive national topics, Chinese models assigned systematically lower ratings to fluency, conciseness, technicality, and overall quality in US-related coverage, while Western models gave higher ratings in the reciprocal China-related evaluation. The sole metric where Chinese-on-US scores exceeded Western-on-China was negative emotion, pointing to a selective amplification of negative affect in politically adversarial contexts.

These patterns align with the hypothesis that large-scale pretraining imbues models with implicit \emph{relational subjectivity}, reflecting not only the statistical structure of language but also the ideological distribution of information in their training data. Prior research on LLM bias has shown that even ostensibly language-neutral embeddings encode political, cultural, and moral priors traceable to their data sources. Our findings extend this literature by showing that such priors manifest not only in explicit classification tasks but also in downstream assessments of news quality—tasks that require integrating multiple linguistic cues.

Importantly, the observed divergences are not random noise but display coherence across semantically related metrics, indicating that they may be stable and replicable features of each model family’s representational space. This has implications for computational journalism, automated content moderation, and cross-cultural NLP applications. In particular, systems that aim to quantify media bias or news quality using LLM-derived features must account for the possibility that their underlying models embed culturally specific evaluative baselines. Without such calibration, cross-source or cross-lingual comparisons risk conflating genuine content differences with model-induced parallax.

Theoretically, these results bridge the conceptual gap between classical media bias frameworks, such as framing theory and agenda-setting, and the emerging paradigm of \emph{algorithmic bias} in AI systems. In both cases, the evaluator—whether human or algorithmic—operates from within a situated epistemic frame, shaped by available information, institutional norms, and implicit value hierarchies. Future work should examine whether these divergences persist across broader topical domains, how they evolve with newer model generations, and to what extent fine-tuning on culturally balanced datasets can attenuate the geopolitical imprint observed here.

\textit{Limitations}
Our analyses depend on probe quality and English-centric corpora; effects may differ for other languages or tasks. Synthetic stress tests do not fully represent real-world harms. Human-label reliability is limited where annotation protocols are absent. We do not claim causal mechanisms; our findings are descriptive diagnostics of embedding behavior.

\section{Conclusions}
This study demonstrates that large language models carry measurable and systematic divergences in their evaluations of news quality, subjectivity, and affective tone, aligned with their geopolitical and cultural origins. By applying a controlled, cross-topic evaluation on politically sensitive domains, we showed that Chinese-origin models tended to emphasize informational novelty and descriptiveness, while Western-origin models assigned higher scores to subjectivity, fluency, and overall presentation quality. The cross-topic asymmetries, particularly the lower structural quality scores from Chinese models on US-related content compared to the reverse, underscore that these divergences are not incidental but patterned.

These findings have two primary implications. First, LLM-based news analysis pipelines are not culturally neutral measurement tools; instead, they reflect the statistical regularities and framing tendencies embedded in their training data. Second, evaluations that ignore the geopolitical provenance of the underlying model risk conflating genuine content differences with model-induced bias, potentially reinforcing or distorting existing narratives. Future work should incorporate cultural calibration, multi-origin ensemble methods, and bias-aware fine-tuning to mitigate these effects. More broadly, our results highlight the need for transparency in LLM selection and for critical awareness when deploying such models in media analysis, journalistic monitoring, or automated content evaluation.


\begin{thebibliography}{00}
\bibitem{lippman}
W. Lippman, \textit{Public opinion}. New York: Classic Books America, 2009.
\bibitem{apibias}
W. D. Dean, “Bias and objectivity,” \textit{American Press Institute}, 18-Jul-2017. [Online]. Available: https://www.americanpressinstitute.org/journalism-essentials/bias-objectivity/. [Accessed: 23-Jun-2020].
\bibitem{time}
M. Pressman, “Journalistic Objectivity: Origin, Meaning and Why It Matters,” \textit{Time}, 25-Feb-2019. [Online]. Available: https://time.com/5443351/journalism-objectivity-history/. [Accessed: 23-Jun-2020].
\bibitem{subj}
S. Steensen, “Subjectivity as a Journalistic Ideal,” in \textit{Putting a face on it: individual exposure and subjectivity in journalism}, Oslo: Cappelen Damm Akademisk, 2017, pp. 25–47.
\bibitem{haberbas}
Habermas Jurgen, \textit{The structural transformation of the public sphere: an inquiry into a category of bourgeois society.} Cambridge, Mass: Massachusetts Institute of Technology, 1991.
\bibitem{economist}
“Google rewards reputable reporting, not left-wing politics,” \textit{The Economist}, 08-Jun-2019. [Online]. Available: https://www.economist.com/graphic-detail/2019/06/08/google-rewards-reputable-reporting-not-left-wing-politics. [Accessed: 23-Jun-2020].
\bibitem{arapakis}
Arapakis, I., Peleja, F., Berkant, B., \& Magalhaes, J. (2016, August). Linguistic benchmarks of online news article quality. In Proceedings of the 54th Annual Meeting of the Association for Computational Linguistics (Volume 1: Long Papers) (pp. 1893-1902).
\end{thebibliography}
\end{document}